\documentclass[useAMS,a4paper,usenatbib]{mn2e}
\usepackage{amssymb, amsmath}
\usepackage{color,graphicx}
\usepackage{tikz}
\usepackage{verbatim}
\usepackage{rotating}

\newcommand{\bea}{\begin{eqnarray}}
\newcommand{\eea}{\end{eqnarray}}

\newcommand{\ApJ}{{Astrophys. J.\,}}
\newcommand{\ApJS}{{Astrophys. J. Suppl.\,}}

\newcommand{\etal}{{\it et al.}}
\newcommand{\be}{\begin{equation}}
\newcommand{\ee}{\end{equation}}

\newcommand{\mc}[1]{{\mathcal{#1}}}
\newcommand{\mdl}{\mc{M}}
\newcommand{\mX}{X}

\newcommand{\mD}{\mc{D}}

\newcommand\lsim{\mathrel{\rlap{\lower4pt\hbox{\hskip1pt$\sim$}}
    \raise1pt\hbox{$<$}}}
\newcommand\gsim{\mathrel{\rlap{\lower4pt\hbox{\hskip1pt$\sim$}}
    \raise1pt\hbox{$>$}}}

\newcommand{\like}{{\mc L}}

\newcommand{\dt}{\mathcal{D}}
\newcommand{\lcdm}{$\Lambda$CDM}

\newcommand{\mM}{{\mathcal{M}}}
\newcommand{\params}{\theta}

\newcommand{\EX}{p(d|\mX)}

\newcommand{\Bxl}{B_{X \Lambda}}
\newcommand{\bBxl}{\overline{B}_{X \Lambda}}
\newcommand{\Omk}{\Omega_\kappa}
\newcommand{\data}{d}
\newcommand{\avB}{\langle B_{i\Lambda} \rangle}
\newcommand{\BXL}{B_{\mX\Lambda}}
\newcommand{\RL}{{\mathcal{R}_\Lambda}}

\newcommand{\archive}[1]{\texttt{arXive: #1}}
\newcommand{\MNRAS}{Mon.~Not.~Roy.~Astron.~Soc.}

\begin{document}
\setlength{\unitlength}{1mm}
\title[Should we doubt the cosmological constant?]{Should we doubt the cosmological constant?}
\author[March et al]{M.C. March$^{1}$, G.D.~Starkman$^{2}$, R.~Trotta$^{1}$ and P.M.~Vaudrevange$^{2}$\\
$^{1}$Imperial College London, Astrophysics Group, 
	  Blackett Laboratory, Prince Consort Road, London SW7 2AZ, UK\\
$^{2}$CERCA/ISO \& Department of Physics, Case Western Reserve University, 10900 Euclid Ave, Cleveland, OH 44106, USA
}
\date{\today}
\maketitle
\begin{abstract}
While Bayesian model selection is a useful tool to discriminate between competing cosmological models, it only gives a relative rather than an absolute measure of how good a model is. Bayesian doubt introduces an unknown benchmark model against which the known models are compared, thereby obtaining an absolute measure of model performance in a Bayesian framework.   

We apply this new methodology to the problem of the dark energy equation of state, comparing an absolute upper bound on the Bayesian evidence for a presently unknown dark energy model against a collection of known models including a flat \lcdm~scenario. We find a strong absolute upper bound to the Bayes factor $B$ between the unknown model and~\lcdm, giving $B \lsim 3$. The posterior probability for doubt is found to be less than 6\% (with a 1\% prior doubt) while the probability for \lcdm~rises from an initial 25\% to just over 50\% in light of the data. We conclude that \lcdm~remains a sufficient phenomenological description of currently available observations and that there is little statistical room for model improvement.
\end{abstract}
\maketitle 
%
%
\section{Introduction}
One of the most important questions in cosmology is to identify the fundamental model underpinning the vast amount of observations nowadays available. The so-called ``cosmological concordance model'' is based on the cosmological principle (i.e., that the Universe is isotropic and homogeneous, at least on large enough scales) and on the hot Big Bang scenario, complemented by an inflationary epoch. This remarkably simple model is able to explain with only half a dozen free parameters observations spanning a huge range of time and length scales. Since both a cold dark matter (CDM) and a cosmological constant ($\Lambda$) component are required to fit the data, the concordance model is often referred to as ``the $\Lambda$CDM model''. It is however important to keep in mind that at this stage the \lcdm~model is not a model in the sense attributed to the word by particle physicists, but rather a phenomenological scenario that appears to be able to explain the vast majority of observations with a great economy of free parameters.

In the classical approach to statistics, models (or hypotheses) can never be proved true, only falsified. \cite{Popper}, for example, argued that theories always remain ``infinitely improbable'' regardless of the amount of evidence gathered in their favour. However, in the context of Bayesian inference support can be accrued for a model if the observed data verify predictions made by the model but not by competing models (see~\cite{JaynesBook}). This is the subject of Bayesian model selection (see e.g.~\cite{TrottaSky2008, Trotta2007} for applications to the cosmological context): given a set of competing models, the Bayes factor gives a measure of the relative performance of each model in explaining the data.  This program naturally prefers models that provide a good fit with the fewest number of free parameters,  thus implementing a quantitative version of Occam's razor.

Although Bayesian model selection can identify the best model in a given set of known models, it has no way of indicating whether the absolute quality of the preferred model is high or low. However, it seems desirable to be able to gauge the absolute performance of a model in a Bayesian sense, rather than just its relative performance with respect to  known competitors. In particular, this seems crucial for deciding whether the set of known  models includes the true model. 

The purpose of this paper is to build on the notion of Bayesian doubt introduced by~\cite{StarkmanTrotta2008} to develop and apply a Bayesian technique for model discovery, focusing in particular on the nature of dark energy.
 The structure of this paper is as follows: in section~\ref{sec:doubt} we recall the notion of Bayesian doubt and introduce a new procedure for estimating an upper bound for the Bayes factor in favour of doubt. We next employ this procedure in section~\ref{sec:cosmo}
 to assess the state of our knowledge of the dark energy equation of state,  focusing on the status of the current \lcdm~concordance model. 
 We present our results in terms of the posterior probability for doubt and for \lcdm~in section~\ref{sec:results} and discuss our conclusions in section~\ref{sec:conclusions}.

\section{Bayesian model discovery}
\label{sec:doubt}

In this section we review the concept of Bayesian doubt and explain how this can lead to model discovery. 

\subsection{The notion of Bayesian doubt}
Bayesian doubt, as introduced by~\cite{StarkmanTrotta2008}, is an extension of Bayesian model selection. It seeks to determine a scale quantifying the absolute quality of a model, as opposed to the relative performance of two models, given by their Bayes factor. The key idea of Bayesian doubt is that the general statistical characteristics of what would be recognised as a `good' model are known, even if the specifics of the model are not. 

We begin by introducing a hypothetical unknown model $\mX$ which has the characteristics of what would be considered a good model, to be defined below. This idealized good model then acts as a benchmark against which  known models can be compared using standard Bayesian model selection. Following~ \cite{StarkmanTrotta2008}, we define `doubt', $\mD$, as the posterior probability of this unknown model:
\be
\begin{aligned}
\label{Doubt}
\mathcal{D} \equiv p(\mX|d) &= \frac{p(d|\mX)p(\mX)}{p(d)} \\
&= \left( 1 + \frac{\sum_i p(d|\mM_i)p(\mM_i)}{p(d|\mX)p(\mX)}  \right)^{-1}
\end{aligned}
\ee 
where $\{ \mathcal{M}_i \}$ ($i=1,\dots,N$) is the set of $N$ known models and $d$ are the data. In the above expression, $p(\mX)$ is the prior  probability for the model X, in other words, the prior probability that our list of known models does not contain the true model.
$p(\mM_i)$ is the prior probability of model $\mM_i$ and $p(d|\mM_i)$ is the Bayesian evidence for model $\mM_i$, given by
\be \label{eq:evidence}
p(d|\mM_i)  = \int d\params_i p(d|\params_i, \mM_i) p(\params_i | \mM_i),
\ee 
where $\params_i$ are the parameters of model $\mM_i$.  $p(d| \params_i, \mM_i)$ is the likelihood function for model $\mM_i$,
and $p(\params_i | \mM_i)$ is the prior probability of the parameters of model $\mM_i$.

Once we have chosen the level of prior doubt by defining the value of $p(\mX)$, 
based on a principle of indifference we assume for simplicity that the prior probabilities for the known $N$ models are all equal, i.e. 
\be \label{eq:priors}
p(\mM_i) = \frac{1}{N}\left(1-p(\mX)\right). 
\ee
We single out the \lcdm~model as one of the set of known models, 
and, looking ahead, refer to it as our baseline model.
Therefore it is useful to rewrite Eq.~\eqref{Doubt} as 
\be \label{Doubt2}
\dt = \left( 1 + \frac{\avB}{\BXL}  \left(\frac{1-p(\mX)}{p(\mX)}\right) \right)^{-1},
\ee 
where we have introduced the Bayes factor
\be
B_{ij} \equiv \frac{p(d|\mM_i)}{p(d|\mM_j)}
\ee
and the average Bayes factor between \lcdm~and each of the known models:
\be
\avB \equiv \frac{1}{N} \sum_{j=1}^N B_{j \Lambda}.
\ee
(Note that the sum over models $\mM_j$ includes $j=\Lambda$ and therefore $\avB \geq 1/N$.)

Rather than looking at $\mD$ directly, one can also consider the relative change in doubt $\mathcal{R}$, given by the ratio of posterior to prior doubt:
\be
\label{Rfactor}
\mathcal{R} \equiv \frac{\mathcal{D}}{p(\mX)} = \left( p(\mX) + (1-p(\mX)) \frac{\avB}{\BXL}   \right)^{-1}.
\ee
A necessary condition for doubt to grow ($\mathcal{R} > 1$) is %
\begin{equation}
\label{testL}
\frac{\avB}{\BXL} \ll 1,
\end{equation}
i.e., that the Bayes factor between model $X$ and \lcdm~be much larger than the average Bayes factor between the known models and \lcdm.

However, for \lcdm~to be genuinely doubted it is not sufficient that $\mathcal{R}  > 1$. One has also to require that the probability for \lcdm~itself decreases, i.e., that $p(\Lambda | d) < p(\Lambda)$. Applying again Bayes theorem, one finds that the ratio of the posterior probability for \lcdm~to its prior probability is given by
\be \label{eq:post_Lambda}
\RL \equiv \frac{p(\Lambda | d)}{p(\Lambda)} = \left( (1-p(\mX)) \avB + p(\mX)\BXL  \right)^{-1}.
\ee
Hence to gather genuine doubt against \lcdm~we require that both conditions $\mathcal{R} > 1$ and $\RL < 1$ be fulfilled. 
 
\subsection{Upper bound on the evidence of the unknown model}
\label{sec:doubtdiss}

In order to apply Bayesian doubt to the problem of cosmological model selection, it is necessary to estimate the evidence of the unknown model,  $\EX$.  The approach suggested by~\cite{StarkmanTrotta2008} was to calibrate the value of $\EX$ on simulated data sets from the best among the known models. This has been shown to lead to model discovery for a toy linear model.  However, in the cosmological context it would be very computationally expensive to implement, even given fast algorithms to compute the evidence, such as MultiNest~\citep{Feroz:2008xx} or the Savage-Dickey density ratio~\citep{Trotta2007}.

In this paper, we put forward a different, more economical approach, which aims at computing an absolute upper bound for $\EX$. Since our aim is to investigate the dark energy sector, in the following we focus on the dark energy equation of state, $w(z)$. We cannot, of course, compute the evidence for $\mX$ explicitly since its parameterization of $w(z)$ is unspecified.  Since the unknown model $\mX$ is to provide a benchmark value for the evidence of the known models, it should be designed to provide a good fit to the available data, including cosmic microwave background (CMB), matter power spectrum (mpk) and supernovae type Ia (SNIa) observations. Therefore, the unknown model should have a high degree of flexibility. At the same time, we do not wish to incur the Occam's razor penalty coming from the high number of free parameters usually associated with a very flexible model. This is because we are seeking to build a phenomenological description for $w(z)$ which, if model $\mX$ is to be a `good' model, should arise from an underlying, presently unknown theory with a small number of free parameters. 

In order to have the advantages of a flexible (and therefore well-fitting) unknown model (i.e. low $\chi^2 /$dof), 
without incurring a penalty for having a large number of free parameters, we define the evidence of the unknown model via the upper bound on the Bayes factor between the \lcdm~baseline model and a stand-in model with a very flexible $w(z)$ (as specified in section~\ref{sec:unknown} below). The absolute upper bound on the Bayes factor $\Bxl$ between the unknown model $\mX$ and \lcdm~(denoted by a subscript $\Lambda$)  is given by (see~\cite{GordonTrotta2007} and references therein for details), 
\begin{equation}
\label{upperevi}
\Bxl  < \bBxl = \exp \left( -\frac{1}{2} (\chi^2_{X} - \chi^2_{\Lambda} ) \right).
\end{equation}
We have defined the best-fit chi-squared as minus 2 the log-likelihood at the best-fit point, $\params_i^*$:
\be
\chi^2_i = -2 \ln p(d|\params_i, \mdl_i){{\Big \vert}_{\params_i = \params_i^*}},
\ee  
where $i=X, \Lambda$. 

The bound of Eq.~\eqref{upperevi} arises by putting {\em a posteriori} the prior probability for the parameters of the stand-in model   into a delta-function located at the observed maximum likelihood value, i.e. by replacing $p(\params_X | \mM_X)$  in Eq.~\eqref{eq:evidence} with $\delta(\params_X-\params_X^*)$. While this prior choice has no Bayesian justification (for it is inappropriate to use {\em a posteriori} information to determine the prior), it does lead to an absolute upper bound on the relative evidence between the baseline~\lcdm~model and the unknown model. In order to calculate the absolute bound of Eq.~\eqref{upperevi}, all that is needed is the difference between the best fit log-likelihood (or chi-squared) of the two models, $\Delta \chi^2 \equiv \chi^2_{X} - \chi^2_{\Lambda}$, which can be easily computed. 
Since the \lcdm~model is nested within the unknown model (i.e., the unknown model reverts to \lcdm~for a specific choice of its parameters leading to $w(z) = -1$), it follows that $\Delta \chi^2 \leq 0$. Therefore it is clear that by construction $\bBxl >1$ always, i.e., that our unknown model is always at least as good as \lcdm. 

By inspecting Eq.~\eqref{upperevi}, one might be tempted to think that this upper bound on the Bayes factor merely translates in Bayesian terms the old goodness-of-fit $\chi^2$ test. For if \lcdm~is a ``bad'' model (on whatever scale one wishes to define this), the value of $\chi^2_{\Lambda}$ will be large and thus the Bayes factor in favour of the unknown model will be large, as well. Thus one might think that Eq.~\eqref{upperevi} simply rephrases the well-known rule-of-thumb of $\chi^2/\text{dof} \sim 1$.  However, this is not the case, for the $\chi^2/\text{dof} \sim 1$ rule only applies asymptotically (for $n\rightarrow \infty$ number of data points) and only if the data points are independent, Gaussian distributed. Those conditions are almost invariably {\em not} met in the cosmological context. For instance, it is not even clear how one would define the concept of degrees of freedom for the CMB data, given that the $C_\ell$'s are not independent and are not Gaussian distributed. In the case of SNIa observations, the $\chi^2/\text{dof} \sim 1$ criterion is satisfied for \lcdm~by construction, for the value of the intrinsic dispersion for the SNe is adjusted in such a way to {\em require} this to be the case, see e.g.~\cite{Kowalski2008}. Therefore one cannot meaningfully use this kind of absolute goodness-of-fit tests on such a data set. 

Instead, the upper bound given by Eq.~\eqref{upperevi} does not require any assumption about asymptotic behaviour, nor that the data are Gaussian distributed, nor independent. One only needs to be able to compute the log-likelihood at the best-fit point, including relevant correlations as necessary. 

Finally, the upper bound of Eq.~\eqref{upperevi} could also be computed using the highest best-fit log-likelihood of all the known models, at no extra computational cost. This would give the absolute upper bound achievable among the class of known models. Although we do not pursue this approach in this paper, we recommend including in any Bayesian model comparison a model $\mX$ with evidence obtained via this procedure, for this will give an  estimate of the maximum possible level of doubt  that can arise from the known models with their assigned priors. 

\subsection{Behaviour of doubt and posterior probability for \lcdm}

\begin{table}
 \begin{tabular}{l l  l} 
  $|\ln B_{ij}|$ & Odds & Strength of evidence \\\hline
 $<1.0$ & $\lsim 3:1$  & Inconclusive \\
 $1.0$ & $\sim 3:1$ & Weak evidence \\
 $2.5$ & $\sim 12:1$  & Moderate evidence \\
 $5.0$ & $\sim 150:1$ & Strong evidence \\
 \hline
\end{tabular}
\caption{Empirical scale for evaluating the strength of evidence from the Bayes factor $B_{ij}$ between two models (so--called
`Jeffreys' scale'). The right--most column gives our
convention for denoting the different levels of evidence above
these thresholds, following~\protect\cite{GordonTrotta2007}.\label{tab:jeff} }
\end{table}

In the following, we will adopt the absolute upper bound $\bBxl$ of Eq.~\eqref{upperevi} as an estimator for the Bayes factor of the unknown model $X$, and explore the consequence in terms of doubt and in terms of the posterior probability for \lcdm. It is clear from Eqs.~\eqref{Doubt2} and \eqref{eq:post_Lambda} that for a given level of prior doubt $p(\mX)$, the posterior models' probabilities are controlled uniquely by the two quantities $\avB$ and $\bBxl$. The result can be expected to fall within one of the three scenarios below, which we will examine from two points of view: using doubt $\dt$ and using the upper bound to the Bayes factor $\bBxl$ as measures of doubt. While there is something to be said for employing $\bBxl$ (whose value can be translated into a strength of evidence via the Jeffreys' scale, given in Table~\ref{tab:jeff}) as a criterion for goodness of fit, it turns out that doubt can shed some light onto how large $\bBxl$ should be to have genuine doubt without referring to the (in some sense arbitrary calibrated) Jeffreys' scale.
\begin{itemize}
\item {\bf Case 1: } $\bBxl \gg 1$ and $\avB \sim 1$: in this case, the unknown model has a much better evidence than \lcdm, which in turn has about the same evidence as the other known models. As the Bayes factor $\bBxl>1$, we should expect there to be a significant amount of doubt, $\dt\approx 1$. And indeed, from Eq.~\eqref{Doubt} the doubt is, assuming $p(\mX) \ll 1$: 
\be
\dt \approx \left(1 + \frac{1}{p(\mX) \bBxl } \right)^{-1}\approx1,
\ee
for $p(\mX)\bBxl \gg1$. In other words, we are inclined to believe that there is a better model that we have not yet thought of if the Bayes factor between the unknown model and \lcdm~is sufficiently large to override the smallness of the prior doubt, $\bBxl>1/p(\mX)$ (notice the independence of the Jeffreys' scale). The change in the probability for \lcdm~itself is given by, from Eq.~\eqref{eq:post_Lambda}, 
\be \label{eq:post_Lambda_1}
\RL \approx \left(1 + p(X) \bBxl\right)^{-1}.
\ee
While the doubt grows ($\dt \rightarrow 1$) the probability for \lcdm~declines, $\RL \ll 1$. In this case, one is led to genuinely doubt \lcdm.
\item {\bf Case 2:} $\bBxl \gg 1$ and $B_{i \Lambda} \ll 1 (i \neq \Lambda)$: in this case, \lcdm~is clearly the best of the known models, as the Bayes factors between the known models and \lcdm~are all small. Again, as the Bayes factor $\bBxl\gg1$ favors the unknown model, we should be doubting our list of models. As $\avB \approx 1/N$, we find
\be
\dt \approx \left(1 + \frac{1}{N p(\mX) \bBxl } \right)^{-1}\approx1.
\ee
for $p(\mX) \bBxl \gg 1/N$. This seems to contradict the result of Case 1. However, as we noted above, the condition that $\dt\approx 1$ is only necessary but not sufficient for doubt to arise. We need to examine the relative change in probability for \lcdm~which is given by
\be \label{eq:post_Lambda_2}
\RL \approx \left(\frac{1}{N} + p(X) \bBxl\right)^{-1}.
\ee
Requiring $\RL < 1$ leads to the stronger condition $p(\mX) \bBxl  \gg 1$, as in Case 1 above. If the latter condition is not fulfilled, doubt will grow at the expenses of the probability of the other known models, as the prior probability mass which was spread among the $N$ known models according to Eq.~\eqref{eq:priors} gets redistributed between $\mX$ and \lcdm. 
\item {\bf Case 3:} $\bBxl \sim 1$: in this case, the upper bound on the Bayes factor between the unknown model and \lcdm~is of order unity. This means that we should have no reason to doubt our set of models. The expression for doubt Eq.~\eqref{Doubt} simplifies to
\be
\dt \approx \left(1 +  \frac{\avB}{p(\mX)} \right)^{-1}.
\ee
In order to reach a high level of doubt $\dt\approx1$, we would need $\avB/p(\mX)\approx0$.
Clearly, this is only the case if we allow for $p(\mX)\gg \avB \leq 1/N$, i.e.~if we are starting off with a prior doubt which is larger than the indifference prior on the known models, which is usually not the case. Otherwise, if the Bayes factor $\bBxl$ is larger than the prior doubt $p(\mX)$, we can regard our list of models as reasonably complete, and perform Bayesian model comparison among the list of known models. Of course, this procedure must be repeated once new data arrives (see~\cite{Starkman:2009xp} for the procedure that this entails). Note that again we do not need to refer to the Jeffreys' scale, but need to compare the average Bayes factor $\avB$ with our prior doubt $p(\mX)$.
\end{itemize}

In summary, we are led to doubt the current baseline \lcdm~model only if the rule of thumb 
\be \label{eq:rule_of_thumb}
p(\mX) \bBxl \gg 1\
\ee
is satisfied, which corresponds to either Case 1 above or to Case 2 when the condition for $\RL < 1$ is also fulfilled. If Eq.~\eqref{eq:rule_of_thumb} is satisfied, we are guaranteed that doubt will grow and at the same time the probability for the \lcdm~model will decrease, thus signaling the opportunity for model discovery. All this is accomplished without referring to Jeffreys' scale.

\section{Application of doubt to the dark energy equation of state}
\label{sec:cosmo}

\subsection{The known models}

We take the flat \lcdm~model as our baseline model, described by the usual 6-parameters set $\params = \{A_s, n_S, \omega_b, \omega_c, \Omega_\Lambda, H_0\}$, where $A_s$ is the amplitude of scalar fluctuations, $n_S$ is the spectral index, $\omega_b$ the physical baryon density, $\omega_c$ the cold dark matter density, $\Omega_\Lambda$ the density parameter for the cosmological constant and $H_0$ the Hubble constant today. We assume purely adiabatic fluctuations throughout this paper. 

We define the other models in the known models list by increasing the complexity of the baseline model in successive steps. First, we only add a non-zero curvature parameter, $\Omk \neq 0$, with a flat prior in the range $-0.3 \leq \Omk \leq 0.3$, akin to the ``Astronomer's prior'' adopted and justified in \cite{Vardanyan:2009ft}. Alternatively, another model is obtained by only adding an effective equation of state parameter for dark energy, $w \neq -1$, with a flat prior in the range $-1.3 \leq w \leq -0.7$ while keeping $\Omk=0$ fixed. Finally, a fourth model with 8 free parameters is obtained by adding both $\Omk \neq 0$ and  $w \neq -1$ with the above priors to the \lcdm~baseline model. 

One could in principle further increase the complexity of the known models, e.g. by adopting more complex descriptions for $w(z)$, such as the so--called CPL parameterization in terms of the parameters $(w_0, w_a)$. However, those models have in general a lower evidence than \lcdm, as they are penalized for their wasted parameter space, see e.g.  
\cite{Liddle:2006kn}. As a consequence, they are expected not to contribute significantly to $\avB$, and therefore their influence on posterior doubt would be minor, see section~\ref{sec:moremodels} for details. One could also add to the list alternative explanations for the apparent acceleration of the Universe, such as for example modified gravity models, provided one can compute their evidence numerically~\citep{Heavens:2007ka}. As the main goal of this paper is to introduce the methodology related to Bayesian doubt, we however restrict our considerations to the four models listed above. We comment in section~\ref{sec:moremodels} on how our results would change if the list of known models would be further enlarged. 

Finally, in this work we do not address the problem of the fine tuning of the value of the cosmological constant itself. All models we consider here suffer equally from the fine tuning problem, i.e., the fact that the measured value of the cosmological constant is some 120 orders of magnitude smaller than the ``natural'' scale set by the Planck mass if $\Lambda$ arises from quantum fluctuations of the vacuum. Anthropic reasoning in the context of the Multiverse has been invoked to explain the smallness of the cosmological constant, and while Bayesian reasoning could be brought to bear on the effectiveness of such an ``explanation'', we shall not consider this aspect further in the present paper. 

\subsection{Parameterization of the unknown model}
\label{sec:unknown}

Our discussion so far has been completely general, sidestepping the crucial issue of how to evaluate Eq.~\eqref{upperevi} for the unknown model. In order to make further progress, we have to make some assumptions regarding the class of alternative models the unknown model $\mX$ is supposed to come from. 

As we are interested in the dark energy sector, we will assume that the phenomenology of model $\mX$ is such that it only leads to modifications to the right-hand-side of Einstein equations. In other words, we do not investigate models that modify General Relativity except for those whose only impact is a change in the effective energy-momentum tensor. Under this assumption, a model $\mX$ is fully specified once we give its redshift-dependent
equation of state of dark energy $w(z)$. Notice that we also implicitly assume that the Universe is well described by a FRW isotropic cosmology. If one wished to include a more general class of alternative models from which to draw $\mX$, one could do so by parameterizing their phenomenology in a suitable way. One could define even more general classes of alternative models, for example by fitting parameterized functions to the observations. However, we do not pursue this approach here, because such a modeling of the data would be devoid of any physical insight and would achieve a purely descriptive fit to the observations. To see why this is not desirable, one only has to push this approach to its extreme consequences: given any data collection, there is always a ``model'' that fits the data perfectly. This model is obtained by simply choosing the value of the ``theory'' to be identical to the observed value for each of the observations. Of course, nobody would ever consider such a model to be a valid scientific theory, because we demand that the latter should have explanatory power, not be a simple description of the data. Therefore, it seems sensible to require from the outset that our unknown model $\mX$ be part of a class of {\em physical} theories, with phenomenological parameters that are linked with the physical framework of the class of models considered (here, FRW isotropic Universes with time-varying dark energy equation of state and otherwise standard cosmology).  

Therefore we are left with the task of parameterizing $w(z)$ as a
function of redshift, and then use its functional form to compute the
$\Delta \chi^2$ between the unknown model and the \lcdm~baseline
model. To this purpose, we employ the Parameterized Post Friedman
(PPF) prescription developed by \cite{Hu:2007pj,
  Hu:2008zd}. The PPF prescription
was originally introduced to describe the behavior of theories of
modified gravity in a metric framework that describes leading order
deviations from general relativity (subject to certain
assumptions). However, it was also found be well-suited for describing
the evolution of dark energy models that cross the so--called
``phantom divide'', $w=-1$. Crossing this phantom divide in models
with fixed sound speed would lead to divergences in the pressure
perturbations. Hence models that are phenomenologically described by a
time-varying $w(z)$ that crosses $w=-1$ must be described
micro-physically by a theory of scalar-fields with a varying speed of
sound, e.g. DGP-type models.  
\subsection{Numerical implementation and data sets}

Below, we investigate the behaviour of doubt for different combinations of cosmological data sets. In particular, we are interested in studying doubt as the constraining power of the combined data increases.

We modified the CosmoMC
\citep{LewisBridle2002}) parameter estimation package to sample the
additional parameters $w_i\equiv w(z_i)$, where $z_i$ are uniformly
spaced at $n=10$ redshift value, ranging from $z=0\dots 1.5$.   \cite{Fang:2008kc, Fang:2008sn} wrote a plugin to
CAMB \citep{LewisChallinor2000} that implements the PPF prescription
and is freely available for
download\footnote{\tt{http://camb.info/ppf/}}, which we adopted for this work. The PPF
module uses cubic splines to interpolate $w$ between these points, and
assumes $w(z>1.5)\equiv w(z=1.5)$.

We adopted the 307 SNe Ia from the ``Union'' data set compiled by \cite{Kowalski2008}. The CMB data and likelihood used was the WMAP five year data set \citep{DunkleyKomatsu2008}. \cite{TegmarkEisenstein2006} provided the data and likelihood code for the matter power spectrum using SDSS DR4. The evidence for the known models is computed using the publicly available MultiNest code~\citep{Feroz:2007kg,Feroz:2008xx,Trotta:2008bp}, which implements the nested sampling algorithm, employed as an add-in sampler to CosmoMC \citep{LewisBridle2002} and CAMB \citep{LewisChallinor2000}. 
 
The gist of nested sampling is that the multi--dimensional evidence integral of Eq.~\eqref{eq:evidence} is recast into a one--dimensional integral. This is accomplished by defining the
prior volume $x$ as ${\rm d} x \equiv p(\params){\rm d}
\params $ so that
 \begin{equation} \label{eq:def_prior_volume}
  x(\lambda) = \int_{\like(\params)>\lambda} p(\params) {\rm d}
  \params
 \end{equation}
where the integral is over the parameter space
enclosed by the iso-likelihood contour $\like(\params) =
\lambda$. So $x(\lambda)$ gives the volume of parameter space
above a certain level $\lambda$ of the likelihood. Then the
Bayesian evidence, Eq.~\eqref{eq:evidence}, can be written
as
 \begin{equation} \label{eq:nested_integral}
 p(\data) = \int_0^1 \like(x) {\rm d} x,
 \end{equation}
where $\like(x)$ is the inverse of
Eq.~\eqref{eq:def_prior_volume}. Samples from $\like(X)$ can be
obtained by drawing uniformly samples from the likelihood volume
within the iso--contour surface defined by $\lambda$. The
1--dimensional integral of Eq.~\eqref{eq:nested_integral} can be
obtained by simple quadrature, thus
 \begin{equation}
 p(\data) \approx \sum_i \like(x_i) W_i ,
 \end{equation}
where the weights are $W_i = \frac{1}{2}(x_{i-1} - x_{i+1})$. The standard deviation on the value of the log-evidence can be estimated as $(H/n_\text{live})^{1/2}$, where $H$ is the negative relative entropy and $n_\text{live}$ is the number of live points adopted, which in our case is $n_\text{live} = 4000$ (see~\cite{Feroz:2007kg} for details). 

The best-fit $\chi^2$ required to evaluate Eq.~\eqref{upperevi} is obtained by performing a Metropolis--Hastings Markov Chain Monte Carlo (MCMC) reconstruction of the posterior of the 16 parameters model comprising the \lcdm~parameters $\theta$ and the above 10-parameters description of $w(z)$. We gather a total of $5\times10^5$ samples in 8 parallel chains and verify that the Gelman \& Rubin mixing criterion~\citep{Gelman92} is satisfied (i.e., $R \ll 0.1$, where $R$ is the inter-chain variance divided by the intra-chain variance). 

MCMC is rather geared towards exploring the bulk of the posterior probability density, and is not particularly optimised to look for the absolute best-fit value. This is especially true for high dimensional parameter spaces. Therefore, we expect that the best-fit $\chi^2$ values recovered via MCMC for the 16-dimensional model $\mX$ are going to be systematically higher than the true best-fit. In order to estimate and correct for this numerical bias, we sampled via MCMC a 16 dimensional Gaussian of unit variance, recovered the best-fit $\chi^2$ and compared it with the true best-fit value, repeating the procedure 5000 times. This gives an estimate of the numerical bias, under the assumption (which is valid locally) that the posterior distribution of model $\mX$ is close to Gaussian in the immediate vicinity of the best-fit. We found that the MCMC systematically overestimates the best-fit $\chi^2$ value by $0.94\pm0.14$, and therefore subtracted this estimate from the recovered $\chi^2$ best-fit value for model $\mX$. We also verified that the numerical bias in recovering the best-fit $\chi^2$ for a 6-dimensional parameter space (such as \lcdm) is negligible in comparison.

\section{Results and discussion}
\label{sec:results}

We now proceed to evaluate the doubt and the posterior probability of \lcdm~for various combinations of cosmological data sets. 

\subsection{Model comparison outcome including doubt}

 In Table~\ref{tab:evid}, we present the estimated upper limit on the Bayes factor between \lcdm~and model $\mX$
 as well as   the Bayes factors with respect to \lcdm~for the other known models. 
 Among the known models, we confirm what many others have shown -- 
 that \lcdm~is the best-fit known model, or at least that no other model is demonstrably better.
Thus, we  find an inconclusive model comparison result (according to the Jeffreys' scale, Table~\ref{tab:jeff})   
when comparing \lcdm~and a model with a free (but constant) $w$. 
We also find weak to moderate evidence  ($1 \lsim \ln B \lsim 2.5$) against spatially curved models when compared to a flat \lcdm, in agreement with the more detailed findings of~\cite{Vardanyan:2009ft}.  Finally, there is weak to moderate evidence against the most complex of the known models -- one exhibiting both $w\neq -1$ and $\Omk \neq 0$. This is in good agreement with the results of previous more thorough analyses, e.g.~\cite{Liddle:2006tc,Liddle:2006kn} and ~\cite{Li:2009}. From this, ordinary Bayesian model comparison concludes that \lcdm~is still the best of the known models (at least for the limited range of alternative models considered here). 

Most importantly, in the table, we report the improvement in the best-fit log-likelihood obtained over \lcdm~by using $\mX$, and use this to compute an absolute upper bound to the Bayes factor via Eq.~\eqref{upperevi}. We notice that the improvement in the best-fit is fairly modest for all the data sets considered, supportive of the general sentiment in the community that \lcdm~is in good agreement with available observations and that therefore there is little room for statistical improvement of the quality of fit.  This is in part because it is very hard to improve the quality of fit by changing $w(z)$ -- observables are usually a double integral of $w(z)$, and therefore insensitive to  features in the equation of state (see e.g~\cite{Huterer:1999};~\cite{Maor:2001} and \cite{Clarkson:2009jq}). As a consequence, even a highly flexible $w(z)$ model such as the one we used here to describe $\mX$ will lead to only small observable departures from the standard cosmological constant scenario.   It is important to keep in mind that such statements depend strongly on the statistics one employs to examine the models.  For example, the standard likelihood function for CMB data is insensitive to most of the reported anomalies in the low-$\ell$ CMB \cite{Copi:2010na,Bennett:2010jb}.

The interesting consequence from the point of view of doubt is that this translates into strong upper limits for the Bayes factor between model $\mX$ and \lcdm~(third from last column of Table~\ref{tab:evid}). We find that the upper limit on the Bayes factor  $\bBxl$ 
(last column of Table~\ref{tab:evid}) for all the data combinations is 
less than $3$, just around ``weak evidence'' threshold ($\ln B = 1$, see Table~\ref{tab:jeff}). 
From our discussion in section~\ref{sec:doubtdiss}, this means that the necessary condition for doubt to grow, $p(\mX) \bBxl \gg 1$, is not met for any reasonable doubt prior choice.  We remind the reader at this point that our unknown model $\mX$ has been designed in such a way as to exhibit the maximum possible evidence against \lcdm. Therefore, if even such a model cannot achieve a significant level of evidence against \lcdm, one can safely conclude that no other reasonable model will.  Of course this conclusion depends both on  the set of observations we have considered and on the particular likelihood function we have ascribed to that data.  New statistical treatments can bring to light anomalies in the existing data, while new observations might contain new unexpected features.
\begin{table*} 
\begin{tabular}{l | c c c | c c | c c }
\hline\hline
               & $-1.3<w<-0.7$  & $w=-1.0$ & $-1.3<w<-0.7$ &  \multicolumn{2}{c}{``Unknown'' model $X$} & $\avB$ & $\bBxl$\\ 
               &  $\Omega_{\kappa}=0.0$ & $-0.3 < \Omega_{\kappa} <0.3$ & $-0.3 < \Omega_{\kappa} <0.3$  \\ 
               &$\ln B_{j\Lambda}$    &$\ln B_{j\Lambda}$&$\ln B_{j\Lambda}$  &$\Delta\chi^2$  & $\ln \bBxl$   \\ \hline
CMB only       &$ 0.18 \pm 0.09$      &$-1.03 \pm 0.09$   &    $-1.09 \pm 0.09$ &$ -1.21 \pm 0.3$&$ 0.61  \pm 0.2$ & $0.72 \pm 0.03$ &$ 1.83 \pm  0.3$\\
CMB + SN       &$-0.37 \pm 0.09$      &$-1.30 \pm 0.09$   &    $-1.63 \pm 0.09$ &$ -2.34 \pm 0.3$&$ 1.17  \pm 0.2$ & $0.54 \pm 0.02$ &$ 3.22 \pm  0.5$\\
CMB + mpk      &$-0.50 \pm 0.08$      &$-2.57 \pm 0.08$   &    $-2.69 \pm 0.08$ &$  -0.88\pm 0.3$&$ 0.44  \pm 0.2$ & $0.44 \pm 0.01$ &$ 1.55 \pm  0.2$\\
CMB + SN + mpk &$-0.48 \pm 0.09$      &$-2.51 \pm 0.09$   &    $-2.73 \pm 0.09$ &$ -2.15 \pm 0.3$&$ 1.08  \pm 0.2$ & $0.44 \pm 0.01$ &$ 2.93 \pm  0.6$\\ \hline
\end{tabular}
\caption{In the first three columns, we report the Bayes factors between the known models and \lcdm~for different combinations of data sets, where $\ln B_{j\Lambda}<0$ favours \lcdm. The fourth columns gives $\Delta\chi^2 = \chi^2_\mX - \chi^2_\Lambda$, the improvement in the best-fit log-likelihood obtained by using model $\mX$ (specified in the text) over \lcdm. The last column gives the corresponding absolute upper bound to the Bayes factor between model $\mX$ and \lcdm.}
\label{tab:evid}
\end{table*}   

Our results in terms of posterior probability for doubt and for the \lcdm~model are shown in Table~\ref{tab:unknownflat}, for two different assumptions regarding the level of prior doubt, $p(\mX) = 10^{-2}$ and $p(\mX) = 10^{-6}$. These two choices are representative of a range that we think might bracket reasonable prior expectations: a prior doubt of 1\% is certainly not too large, while leaving a little space for updating our models beliefs in the light of data. A prior doubt of $10^{-6}$ reflects the fact that surely we have to allow for a one-in-a-million chance that our current list of known models might be incomplete, and that the true underlying dark energy model might still be undiscovered. 

\begin{figure}
\centering
\includegraphics[width=\linewidth]{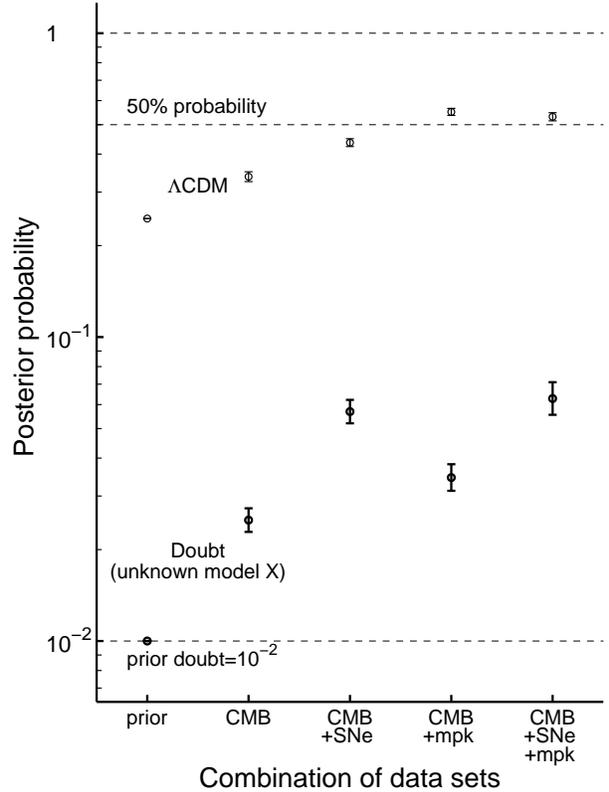}
\caption{Posterior probability for doubt for the \lcdm~model as a function of different combinations of data sets. The probability of \lcdm~increases from the initial 25\% to just about over 50\%, while the probability of doubt increases from the initial 1\% to just over 6\%, mostly as a consequence of acquiring probability from the other 3 known models considered in the analysis. This signals that \lcdm~remains the most valid statistical description of the data.}
\label{fig:doubt}
\end{figure}

Table~\ref{tab:unknownflat} contains the level of doubt, which is updated from the prior by using the results of Table~\ref{tab:evid} for the models' evidences. We find an increase in doubt by a factor of $\sim 6$ for the most constraining data combination (CMB+SN+mpk). This however is largely a consequence of the doubt model acquiring some of the probability mass of the known models other than \lcdm, as discussed under Case 3 in section~\ref{sec:doubtdiss}. Indeed, the posterior probability of \lcdm~is observed to increase (last column of Table~\ref{tab:unknownflat}), from the initial prior value $p(\Lambda) \approx 0.25$ to just over 50\% for the most constraining data combination. This result is almost independent of the choice of prior doubt. The behaviour of the posterior probability for doubt and \lcdm~for a prior choice $p(\mX) = 10^{-2}$ is shown in Fig.~\ref{fig:doubt}, as a function of the data sets employed.

\begin{table*} 
\begin{tabular}{lcccc}
\hline\hline
&\multicolumn{2}{c} {Doubt $\dt$} & Posterior for \lcdm, $p(\Lambda | d)$     \\
 & Prior doubt: $p(\mX)= 10^{-2}$&
Prior doubt: $p(\mX)=10^{-6}$ & (with $p(\mX) = 10^{-2}$ and $p(\Lambda) \approx 0.25$) \\ \hline
CMB only  &$(2.50 \pm0.2)\times10^{-2}$&$(2.54\pm0.2)\times10^{-6}$ & $0.34 \pm 0.01$\\ 
CMB+SN    &$(5.69 \pm0.5)\times10^{-2}$&$(5.97\pm0.6)\times10^{-6}$ & $0.44 \pm 0.01$\\ 
CMB+mpk   &$(3.46 \pm0.3)\times10^{-2}$&$(3.54\pm0.4)\times10^{-6}$ & $0.55 \pm 0.02$\\ 
CMB+SN+mpk&$(6.29 \pm0.8)\times10^{-2}$&$(6.64\pm0.9)\times10^{-6}$ & $0.53 \pm 0.02$ \\       \hline
\end{tabular}
\normalsize
\caption{First two columns: Posterior doubt for different data sets combinations and two prior doubt assumptions. Last column: posterior probability for the \lcdm~model when allowing for the possibility of a 1\% prior doubt on the completeness of our list of known models. }
\label{tab:unknownflat}
\end{table*}   

\subsection{Impact of the addition of further known models}
\label{sec:moremodels}

\begin{table}
 \begin{tabular}{l l  l l}
Known models    & \multicolumn{3}{c}{Required $\Delta\chi^2$ for $p(\mX | d) = p(\Lambda | d)$} \\  
  $N$ & $p(\mX) = 10^{-2}$ & $f$ & $p(\mX) = p(\Lambda)/f$  \\\hline
  4 & $-6.4$  & 4 & $-5.5$\\ 
  10 & $-4.6$ & 10 & $-9.2 $ \\
  20  & $-3.2$ & $10^2$ & $-18.4$\\
  50  & $-1.4$ & $10^3$ & $-27.6$ \\ 
 \hline
\end{tabular}
\caption{Improvement in the $\chi^2$ of \lcdm~required for the unknown model $\mX$ to have the same {\em a posteriori} probability as \lcdm. First two columns: as a function of the number of known models, $N$, assuming a fixed prior doubt $p(\mX) = 10^{-2}$. Last two columns: assuming a fixed fractional prior doubt, $p(\mX) = p(\Lambda)/f$, and as a function of $f$.  It is assumed that the evidence of the known models is much smaller than the evidence for \lcdm. \label{tab:chisq} }
\end{table}

\begin{figure}
\centering
\includegraphics[width=\linewidth]{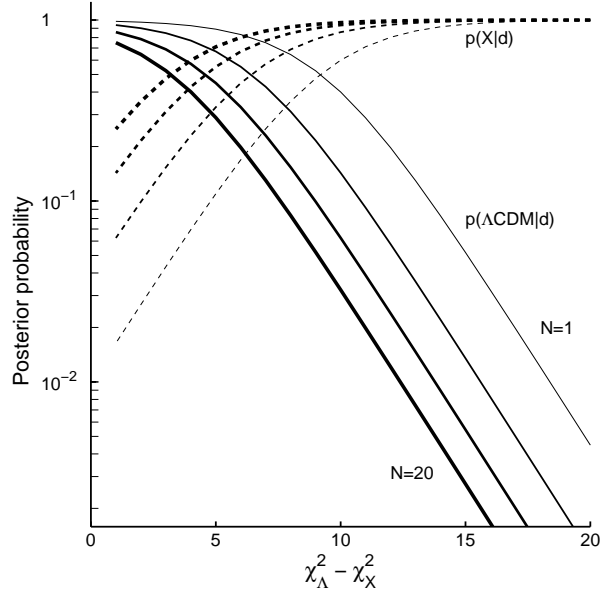}
\caption{Posterior for doubt (dashed lines) and for \lcdm~(solid lines) as a function of $-\Delta\chi^2 = \chi^2_\Lambda - \chi^2_X$ assuming a fixed prior doubt $p(\mX) = 10^{-2}$. Different curves are for different numbers of known models, $N=1,4,10,20$ (from thin to thick), assuming that $\avB \approx 1/N$.}
\label{fig:doubt_behaviour}
\end{figure}

We now proceed to estimate the robustness of our findings with respect to expanding the set of known models.  As has been mentioned above, the list of three alternative known models to \lcdm~we adopted in this work is far from complete. However, even if a larger number of models $N$ were included in the known models list, it is reasonable to assume that the value of the average evidence between the known models and \lcdm~would scale approximately as $\propto 1/N$, for there is no other known model that presently can achieve a substantially higher evidence than \lcdm~(if this was the case, then this other best model would take the place of \lcdm~and become our baseline model which we seek to doubt -- or rather the dominant model in our list of models where we intend to compute the doubt for the whole list). By equating Eqs.~\eqref{Doubt2} and \eqref{eq:post_Lambda} we can solve for the value of $\Delta\chi^2$ required for the posterior on doubt to be equal to the posterior of \lcdm. This gives the approximate condition (assuming that $\avB \approx 1/N$ and that $p(\mX)  \ll 1$):
\be
\Delta\chi^2 \approx 2 \ln (p N).
\ee
So the value of $\Delta\chi^2$ required for posterior doubt to reach the posterior for \lcdm~scales logarithmically with the number of known models. Assuming a prior doubt $p(\mX) = 10^{-2}$ one obtains the values of $\Delta\chi^2$ listed in the first column of Table~\ref{tab:chisq} as a function of $N$. As more known doubts are put on the table, it becomes easier to doubt \lcdm. From this scaling, it would appear that the improvement of $\Delta \chi^2 = -2.3$ for model $\mX$ reported in Table~\ref{tab:evid} for the data combination cmb+SN would lead to a larger probability of doubt than for \lcdm~if we had assumed a list of $N \gsim 30$ known models, rather than just three. As illustrated in Fig.~\ref{fig:doubt_behaviour}, this effect is however a consequence of our choice of spreading the level of prior probabilities among the $N$ known models, while assuming {\em a fixed} $p(\mX)$, see Eq.~\eqref{eq:priors}. As $N$ increases, the prior for \lcdm~decreases while the prior doubt is kept constant. As a consequence, it becomes easier for the former to ``catch up'' with the latter. 

In order to avoid this spurious effect, one could choose to set the prior doubt as a fraction $1/f$ ($f>1$) of the prior probability for \lcdm, i.e., to require that the relative probability between $\mX$ and $\Lambda$ is constant {\em a priori}, independent of the number of known models. We thus replace the prescription of Eq.~\eqref{eq:priors} by 
\begin{align}
p(\Lambda) & = \frac{1}{N}(1-p(\mX)) \\
p(\mX) & = \frac{p(\Lambda)}{f} = (Nf + 1)^{-1}
\end{align}
and by equating the posterior doubt with the posterior for \lcdm~we obtain the following requirement for the $\Delta\chi^2$:
\be
\Delta \chi^2 = -4 \ln f.
\ee
This is now independent of the number of known models $N$ and it only depends logarithmically on the prior doubt fraction, $f$. From the last two columns of Table~\ref{tab:chisq} we can see that even if doubt started off a factor of just $f=4$ less probable than \lcdm, a $\Delta\chi^2 = -5.5$ would be required in order for the unknown model $\mX$ to become as probable as \lcdm. Increasing the prior gap between doubt and \lcdm~(i.e., increasing $f$) only makes the requirements on the $\chi^2$ improvement more taxing. 

In summary, once the effect of adding extra models to the known models' list is corrected for by introducing the fractional prior doubt $f$, we find that the improvement in the $\chi^2$ found for various combinations of data sets is insufficient to doubt \lcdm. If the unknown model starts off being a factor of 4 less probable than \lcdm, one would need an improvement in the $\chi^2$ of about 5 units to reverse the situation in the posterior, which is quite a bit larger than the maximum $\chi^2$ improvement observed from the data. 

\section{Conclusions}
\label{sec:conclusions}

The aim of this paper was to extend the application of Bayesian model selection to define an absolute scale of goodness of fit for models, rather than just a relative one, such as the Jeffreys' scale. We showed how the notion of doubt can be used to evaluate the evidence in favour of a missing `ideal' unknown model in the list of known cosmological models. We demonstrated how a useful absolute upper bound to the Bayesian evidence of an unknown model can be derived and how this can be implemented in the context of Bayesian model comparison. 

Doubt can be incorporated in the framework of model comparison to help us decide whether our currently ``best'' model is statistically adequate for the data at hand. \cite{Kunz:2006mc} introduced the notion of Bayesian complexity to decide whether the available models are over-complex with respect to the constraining power of the data. Bayesian doubt can act as a useful complement to Bayesian complexity, giving an indication of whether the current models are statistically insufficiently to describe the data. Used in conjunction, doubt and complexity can thus extend the power and domain of applicability of Bayesian model comparison. Of course statistical considerations should never replace proper physical insight: all of our arguments are restricted to the statistical aspects of data modeling. But for the problem of dark energy, where most ``models'' are of a phenomenological kind, it seems to us that a rigorous statistical framework can help deciding whether new theoretical explorations might be fruitful. Other domains where we expect doubt to be useful include the description of the spectral distribution of CMB anisotropies and the problem of anomalous alignments between multipoles in the CMB~\citep{Tegmark:2003,Schwarz:2004,Land:2005}.

We have applied this methodology to the problem of dark energy, adopting a list of known models including possible extensions of the dark energy sector and non-zero curvature of the Universe. In principle, many more models could be added to the list of known models. However we argued that our results are robust against adding further models to the list of known models. We found that current CMB, matter power spectrum and SNIa data do not require the introduction of an alternative model to the baseline flat \lcdm~model. The upper bound of the Bayesian evidence for a presently unknown dark energy model against \lcdm~gives only weak evidence in favour of the unknown model. Since this is an absolute upper bound, we conclude that \lcdm~remains a sufficient phenomenological description of currently available observations.

\textit{Acknowledgements.} MCM was partially supported by a travel grant by the Royal
Astronomical Society. MCM would like to thank CWRU for hospitality. PMV would like to thank Imperial College London for hospitality. PMV was supported by NASA ATP grant NNX07AG89G to Case Western Reserve University. GDS was supported by a grant from the US DoE to the theory group at CWRU.
We would like to thank Pietro Berkes, Andrew Jaffe and  Ben Wandelt for useful discussions. Numerical calculations were carried out at Case Western's High Performance Computer Cluster and at the Imperial College High Performance Computing Service. 
We acknowledge the use of the Legacy Archive for Microwave Background Data Analysis (LAMBDA). Support for
LAMBDA is provided by the NASA Office of Space Science.


\end{document}